# Spectroscopy of cold rubidium Rydberg atoms for applications in quantum information

I.I. Ryabtsev, I.I. Beterov, D.B.Tretyakov, V.M. Entin, E.A. Yakshina

*Rzhanov Institute of Semiconductor Physics SB RAS, 630090 Novosibirsk, Russia*
*Novosibirsk State University, 630090 Novosibirsk, Russia*
*E-mail*: ryabtsev@isp.nsc.ru

*Abstract*

Atoms in highly excited (Rydberg) states have a number of unique properties which make them attractive for applications in quantum information. These are large dipole moments, lifetimes and polarizabilities, as well as strong long-range interactions between Rydberg atoms. Experimental methods of laser cooling and precision spectroscopy enable the trapping and manipulation of single Rydberg atoms and applying them for practical implementation of quantum gates over qubits of a quantum computer based on single neutral atoms in optical traps. In this paper, we give a review of the experimental and theoretical work performed by the authors at the Rzhanov Institute of Semiconductor Physics SB RAS and Novosibirsk State University on laser and microwave spectroscopy of cold Rb Rydberg atoms in a magneto-optical trap and on their possible applications in quantum information. We also give a brief review of studies done by other groups in this area.

# 1. Introduction.

During the last decade, significant advances have been made in the laser cooling and trapping of neutral atoms [1-3], both abroad (hundreds of the papers published) and in Russia [4-10]. The most widespread are setups with magneto-optical and optical traps which enable temperatures of translatory atom motion below 100 μK to be reached. Using ultracold trapped atoms opens up new opportunities for various types of basic and applied research in the field of atomic spectroscopy due to spatial localization of atoms in a small volume and the absence of the transit-time and Doppler broadenings. For instance, the time of cold atom interaction with probe radiation is defined by the lifetime of the atom in the trap, which can reach tens of seconds. At the low atomic density, the width of some optical resonances in atoms might be below 1 Hz thus allowing us to consider the atomic frequency standards of a new generation [7]. On the other hand, the resonances should undergo broadenings and shifts at a high atom density due to collisions and interatomic interactions. Therefore, using the methods of laser and microwave spectroscopy, one can investigate the collisional and collective processes in cold atom ensembles [4]. Finally, cold atoms in optical traps are viewed as a promising object to create the qubits of a quantum computer [11].

One of the new areas in atomic spectroscopy became concerned with the experimental study of ultracold atoms in highly excited (Rydberg) states. Rydberg atoms with the principal quantum number $n \gg 1$ represent a unique object for high-resolution spectroscopy and experiments in the field of quantum electrodynamics due to long lifetimes, transition frequencies in the infrared and microwave domains of the spectrum, values of the transition dipole moments on the order of a few thousand atomic units and more, and high sensitivity to external electromagnetic fields [12,13]. These properties enable, for example, exciting one- and multiphoton transitions between Rydberg states at a very low radiation intensity and conducting spectroscopic studies under conditions where spontaneous level relaxation is absent. In addition, for weakly bound electrons in Rydberg atoms, one can build up precise theoretical models of the atomic interaction with electromagnetic field and compare them with experimental data.

A unique feature of Rydberg atoms is the opportunity to detect them with a selective field ionization (SFI) technique by a weak electric field (an electric field with the strength exceeding some critical value, defined by the quantum numbers $nL$, ionizes the atom with probability close to unity), so that using an electron multiplier one can detect single atoms [12-15]. This method provides determination of both the total number of Rydberg atoms and the relative populations of various Rydberg states. In this way it is possible to conduct experiments with a single or a few atoms, which are extremely important, for example, for quantum information processing. The properties specified above allow us to use atoms in Rydberg states for novel topical studies in the field of atomic and quantum physics, where one needs to be able to control the motions and to monitor the states of individual quantum particles.

In this paper, we give a review of the experimental and theoretical studies performed by the authors at the Rzhanov Institute of Semiconductor Physics SB RAS and Novosibirsk State University on laser and microwave spectroscopy of cold Rb Rydberg atoms in a magneto-optical trap and on their possible applications in quantum information processing [4,16-25]. We also give a brief review of the work conducted by other groups on this subject.

## 2. Spectroscopy of the three-photon laser excitation of cold Rb Rydberg atoms in a magneto-optical trap.

Three-photon laser excitation is of interest because it enables the excitation of cold Rydberg atoms without the recoil and Doppler effects (Fig. 1), thus excluding atom heating during photon absorption and providing high fidelity of the quantum gates performed over them. In our theoretical paper [19] we have shown that for such excitation one needs to have a zero sum of wave vectors of the exciting laser radiations. The single-photon $5S \rightarrow nP$ and two-photon $5S{\rightarrow}5P{\rightarrow}nS,nD$ excitation schemes of Rb Rydberg atoms (Fig. 1a) do not satisfy this condition. It can be satisfied in the three-step scheme $5S{\rightarrow}5P{\rightarrow}6S{\rightarrow}nP$ if the three laser beams are incident from three different sides at certain angles with respect to each other (Fig. 1b). The intermediate resonances of the three-photon transition should have large detunings ($\sim$1 GHz) in order not to populate the short-lived intermediate levels. In this case, the calculated excitation spectra of Rydberg states exhibit the narrow Doppler-free resonances with the width defined only by the interaction time of atoms with laser radiation, both for the ultracold and room-temperature atoms (Fig. 1c). We have not, however, effected this star-like geometry yet, because it demands a special magneto-optical trap with certain configuration of the optical windows in its vacuum chamber.

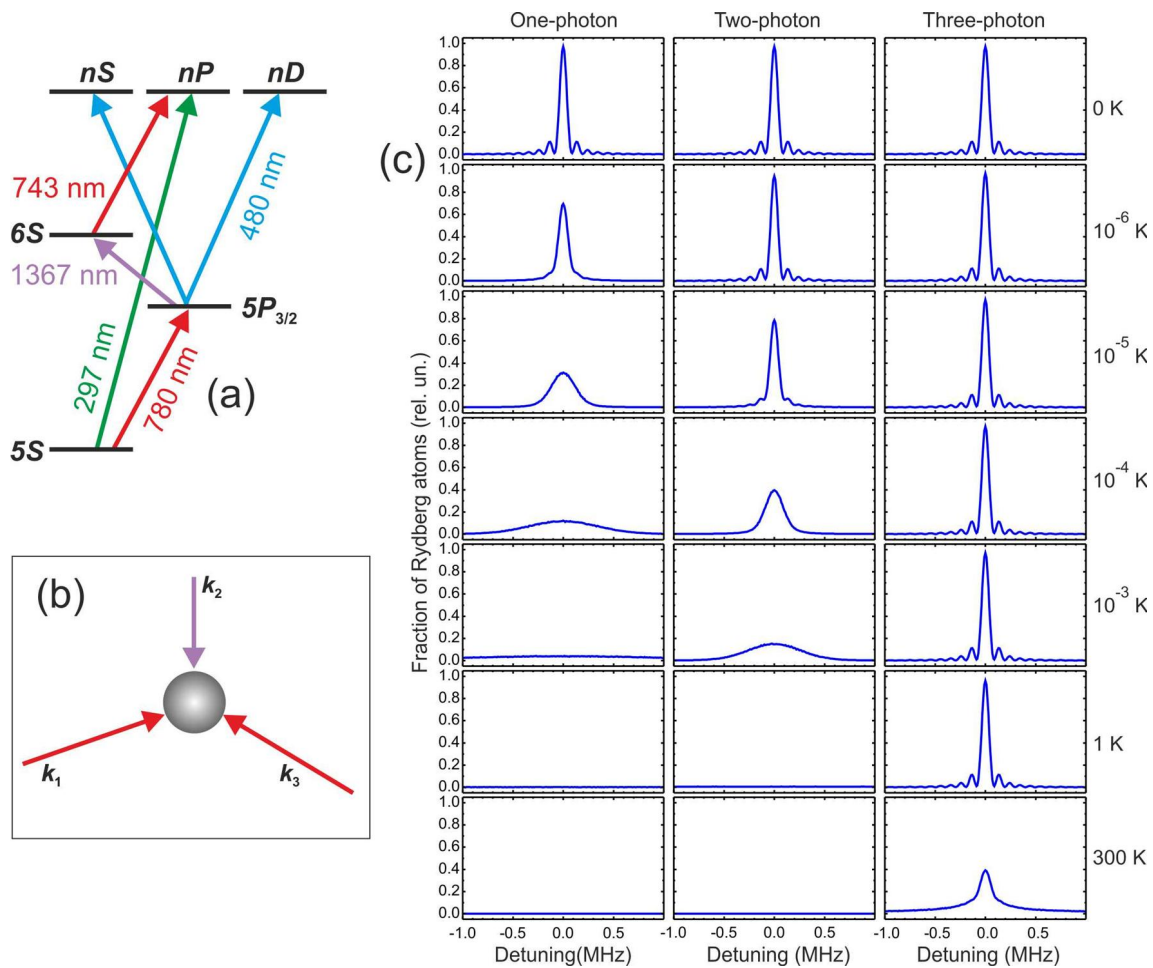

**Fig. 1.** (a) Schemes of one-, two- and three-step excitation of Rydberg states in Rb atoms. (b) Geometry of the three laser beams for Doppler- and recoil-free three-photon excitation. The sum of their wave vectors $\mathbf{k_1}$, $\mathbf{k_2}$ and $\mathbf{k_3}$ should be equal to zero. (c) Calculated excitation spectra of the Rydberg states by a $\pi$ laser pulse of the 10 $\mu$s duration for various temperatures (shown in the right-hand part of the figure) of the atom sample.

We performed our first experiments on precision spectroscopy of the three-photon laser excitation of cold Rb Rydberg atoms in a continuously working magneto-optical trap (MOT) [21]. In these experiments, Rb atoms trapped in a MOT (Fig. 2a) and cooled to a temperature of $\sim$150 $\mu$K were excited to the Rydberg $nP$ states. We implemented for the first time a scheme of three-photon excitation $5S_{1/2} \rightarrow 5P_{3/2} \rightarrow 6S_{1/2} \rightarrow nP$ using cw single-frequency lasers at each excitation step (fig. 2b). At the first step, we applied the radiation of the cooling laser (external-cavity diode laser) with the 780 nm wavelength, which was red detuned by $\delta_1$=-(10-15) MHz from the frequency of the $5S_{1/2}(F{=}3){\rightarrow}5P_{3/2}(F{=}4)$ transition in the $^{85}$Rb isotope.

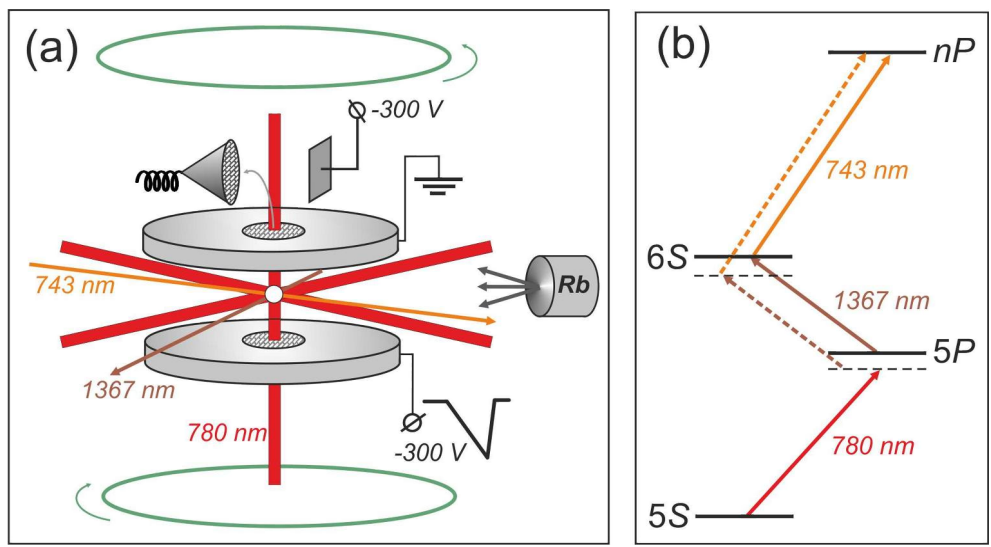

**Fig. 2.** (a) Schematic of the experiment with cold Rb Rydberg atoms in a magneto-optical trap (MOT). Rydberg atoms are excited in a small volume of the cold atom cloud and detected by the selective field ionization technique. (b) Diagram of the three-photon laser excitation $5S_{1/2} \rightarrow 5P_{3/2} \rightarrow 6S_{1/2} \rightarrow nP$ of Rb Rydberg atoms in a continuously operated MOT.

At the second step, we used a telecom distributed-feedback diode laser tuned on exact resonance $\delta_2 = 0$ with the transition $5P_{3/2}(F=4) \rightarrow 6S_{1/2}(F=3)$. Spectra of the three-photon excitation were recorded by scanning the detuning $\delta_3$ of the third-step laser with the 743 nm wavelength (cw ring titanium-sapphire laser). The radiation of the first-step laser was applied continuously, while the radiations of the second- and third-step lasers were supplied either continuously or in a pulsed regime using external modulators, which formed the pulses of the 1-100 µs duration. The radiations of the second and third steps were introduced into the MOT in the geometry of two tightly focused crossed beams, thus providing a localized excitation in a small region with the characteristic size of about 20 µm. The main feature of the experiment was the detection of single Rydberg atoms with the selective field ionization technique with a resolution over the number of detected atoms.

The gradient magnetic field of the MOT was not switched off, but its influence was minimized by adjusting the position of the excitation volume to the point of zero magnetic field. This was monitored via the absence of the Zeeman splitting of the microwave transition $37P_{3/2} \rightarrow 37S_{1/2}$ at 80 GHz frequency using the method described in our paper [16]. This allowed us to maintain a high repetition rate of the laser pulses (5 kHz) and to monitor the changes in the signals from Rydberg atoms in real time on an oscilloscope screen and in a computer-based data acquisition system.

For the continuously operated MOT, when the cooling laser beam (780 nm wavelength) continuously excited the first step with the red detuning $\delta_1 = -(10-15)$ MHz, in the excitation spectra we observed two partly overlapping peaks of different amplitudes, both for the low Rydberg state $37P$ and for the high Rydberg state $77P$ (Fig. 3). The high-frequency peak corresponds to the coherent three-photon excitation without populating the intermediate states, as shown by the dashed arrows in fig. 2b, while the low-frequency peak corresponds to the incoherent tree-step excitation via partly populated intermediate states $5P$ and $6S$ (solid arrows in Fig. 2b). In order to analyze the spectra, we developed a four-level theoretical model based upon the optical Bloch equations [21]. A good agreement between experiment and theory has been obtained by introducing into the theoretical model an additional decay of the optical coherence due to finite laser linewidths (the red curves in Fig. 3). We have shown that other sources of broadening (stray electromagnetic fields, Doppler broadening, interatomic interactions) can be accounted for in this model by increasing the effective linewidth of the third-step laser, as we have done for the 77$P$ state.

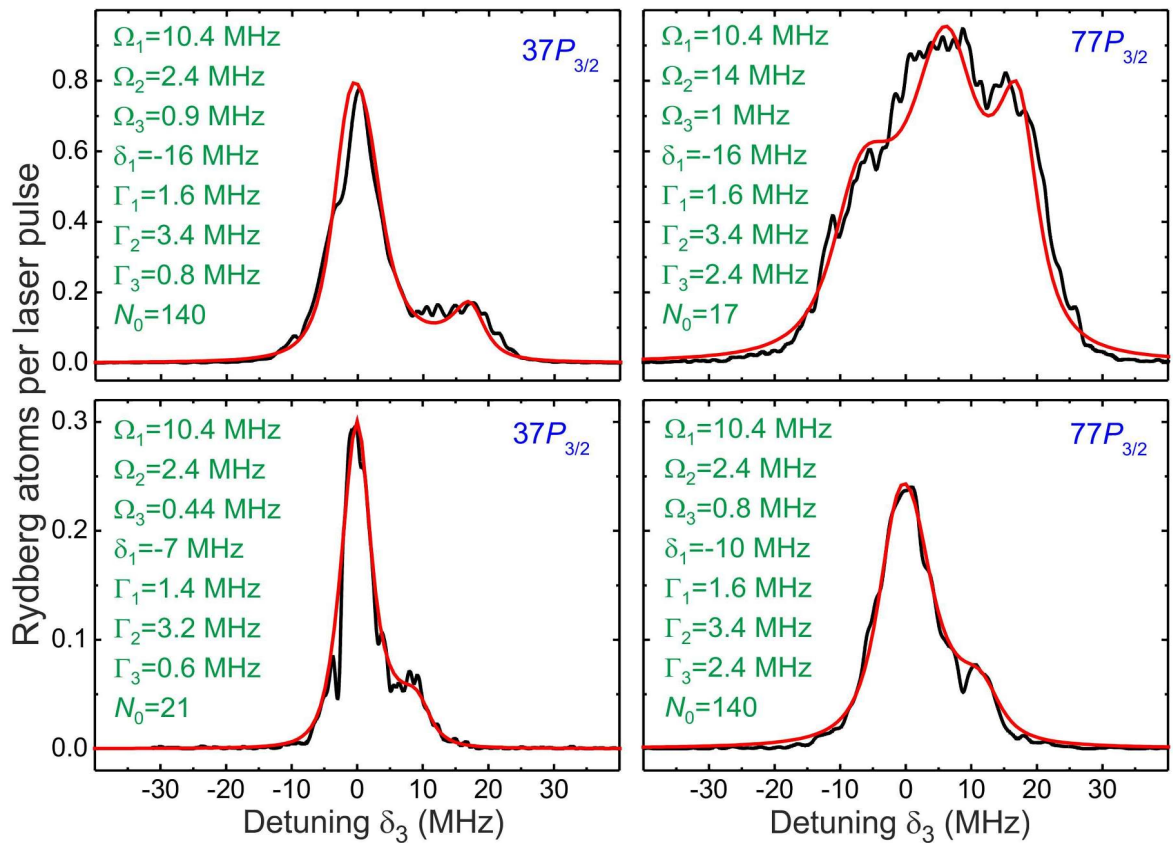

**Fig. 3.** (Color online) Comparison between experimental (black curves) and theoretical (red curves) spectra of the three-photon laser excitation $5S_{1/2} \rightarrow 5P_{3/2} \rightarrow 6S_{1/2} \rightarrow nP$ of cold Rb Rydberg atoms in a continuously operated MOT. Shown are those parameters which have been used in fitting the theoretical model: Rabi frequencies $\Omega_1 - \Omega_3$, detuning of the cooling laser on the first step $\delta_1$, total laser linewidths and other sources of broadenings on each step $\Gamma_1 - \Gamma_3$, and the number of atoms in the excitation volume $N_0$. The interaction time is 4 µs. The detuning of the second-step laser is $\delta_2 = 0$.

We conducted the further experiments on the spectroscopy of three-photon excitation in a MOT which was preliminarily switched off for a short time (20-50 µs) [23]. To this end, all laser beams were equipped with optical modulators that formed the laser pulses of the 1-3 µs duration, while the cooling laser beams were switched off in advance. The first-step laser with the 780 nm wavelength was blue detuned by $\delta_1$=+92 MHz from the transition $5S_{1/2}(F=3) \rightarrow 5P_{3/2}(F=4)$ in order to decrease the population of the intermediate states 5$P$ and 6$S$. The second-step laser with the wavelength 1367 nm was tuned on exact resonance with the transition $5P_{3/2}(F=4) \rightarrow 6S_{1/2}(F=3)$, while the frequency of the third-step laser was scanned across the $6S \rightarrow nP$ transition (Fig. 4a).

In this case, we observed a narrow peak of coherent three-photon excitation detuned by $\delta_3$=-92 MHz at sufficiently large Rabi frequencies of the intermediate single-photon transitions, while the peak of incoherent three-step excitation was suppressed and had a strong power splitting due to ac Stark effect on the 6$S$ state (Autler-Townes effect) as shown in Fig. 4b. On reducing Rabi frequency on the second step, this splitting decreased (Fig. 4c). A comparison with the results of numerical simulations (the red curves in Figs. 4b,c) shows a good agreement between experiment and theory if the finite laser linewidths are taken into account. A small peak at zero detuning appeared due to incomplete switching off of the cooling laser radiation in an acousto-optical modulator. The appearance of this peak allowed us to identify the position of the unperturbed resonance in the spectra of three-photon excitation.

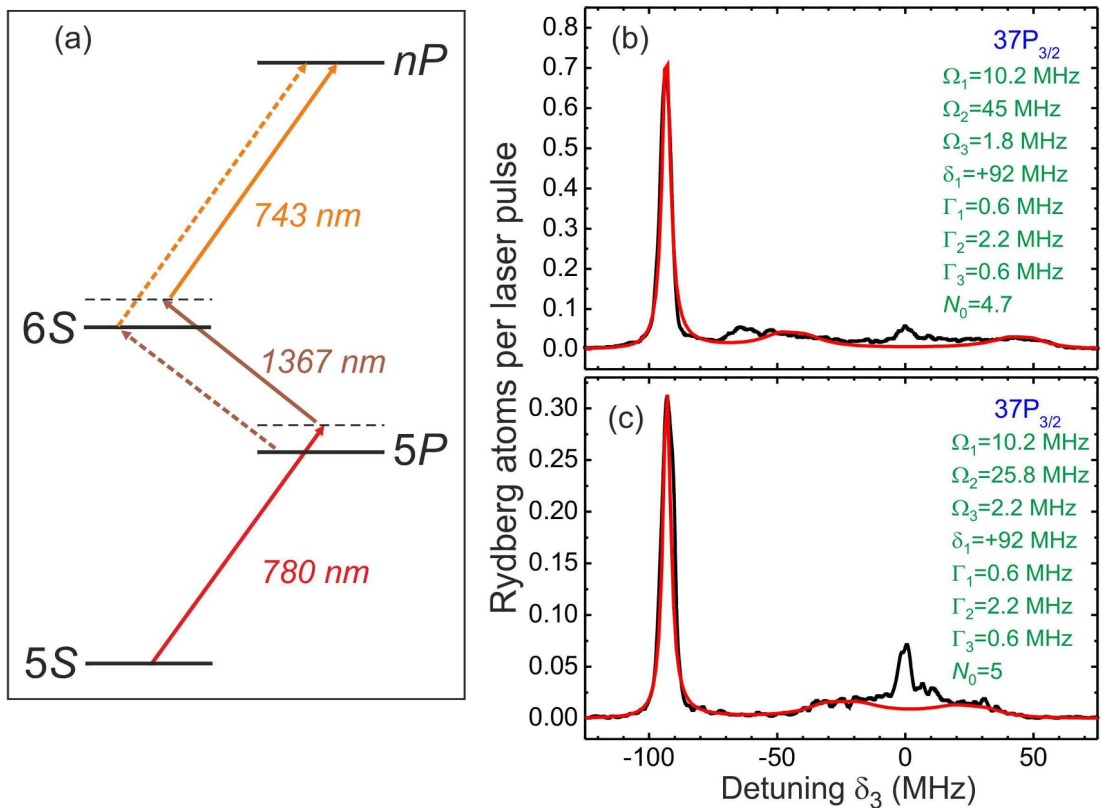

**Fig. 4.** (Color online) (a) Diagram of the three-photon excitation of the Rydberg $nP$ states in Rb atoms in a magneto-optical trap which was switched off. The first-step laser is blue detuned by +92 MHz. (b) Record of the spectrum of the three-photon excitation at a large intensity of the second-step laser. (c) Record of the spectrum of the three-photon excitation at lower intensity of the second-step laser. The red curves depict theoretical calculations. The green labels are the parameters of calculations: Rabi frequencies $\Omega_1 - \Omega_3$, detuning of the laser on the first step $\delta_1$, total laser linewidths and other sources of broadening at each step $\Gamma_1 - \Gamma_3$, and the number of atoms in the excitation volume $N_0$. The interaction time is 2 μs. The detuning of the second-step laser is $\delta_2 = 0$.

## 3. Controlling the interactions of cold Rb Rydberg atoms with Stark-tuned and radio-frequency assisted Förster resonances.

Experimental implementation of two-qubit quantum gates over cold neutral atoms demands that their interactions be switched on and off. This can be realized by short-term excitation to the strongly-interacting Rydberg states [26-29].

The study of the interactions of Rydberg atoms is thus a key factor for creating a quantum computer with qubits on neutral atoms in optical traps. Depending on the particular Rydberg states, these can be either van-der-Waals or resonant dipole-dipole interactions, which feature different dependences on the interatomic distance $R$ ($R^{-6}$ or $R^{-3}$, correspondingly). The dipole-dipole interaction is stronger at long distances (exceeding the size of the Rydberg atom $\sim a_B n^2$, where $a_B$ is the Bohr radius), and its usage is preferred to increase the fidelity of quantum gates performed with Rydberg atoms.

In our first experiments [4,16,20] the interaction between a few cold Rb Rydberg atoms in the 37$P_{3/2}$($|M_J|$=1/2) state, which were positioned in a laser excitation volume of 20-30 μm in size, was controlled by the Förster resonance Rb(37$P$)+Rb(37$P$)→Rb(37$S$)+Rb(38$S$) as shown in

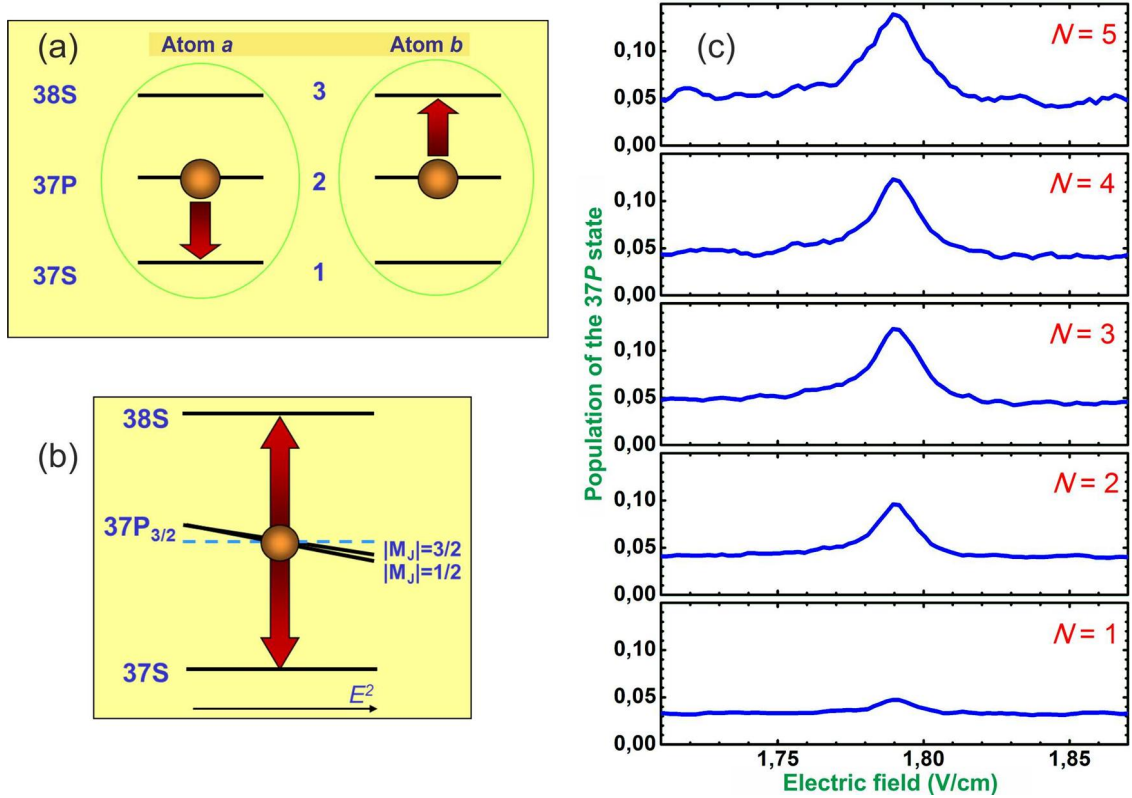

**Fig. 5.** (a) Schematic of the Förster resonance Rb(37$P$)+Rb(37$P$)→Rb(37$S$)+Rb(38$S$) for two cold Rb Rydberg atoms in the 37$P_{3/2}$(|$M_J$|=1/2) state. The atoms undergo resonant dipole-dipole interaction by exchanging virtual photons. (b) Förster resonance arises when the laser-excited Rydberg state 37$P_{3/2}$(|$M_J$|=1/2) lies midway in energy between the neighboring levels 37$S$ and 38$S$. This is achieved using the quadratic Stark effect in a weak electric field $E$. (c) Probability of the transition to the final state 37$S$ for $N$=1-5 Rydberg atoms detected by the selective field ionization technique. The amplitude and width of the resonance grow as the number of atoms rises, in agreement with theory [19].

Fig. 5a. The resonance occurs when the energy of the laser-excited Rydberg level 37$P$ is exactly midway between the neighboring levels 37$S$ and 38$S$, so that the atoms experience resonant dipole-dipole interaction. In the conditions of the Förster resonance, one of the atoms in the 37$P$ state goes down in energy to the 37$S$ state and emits a virtual photon, while another atom absorbs this photon and goes to the higher state 38$S$. Rydberg atoms experience the van-der-Waals interaction beyond the exact energy resonance. The resonance is obtained by applying a dc electric field 1.79 V cm$^{-1}$ (Fig. 5b), and the measured value is the probability of transition to the final 37$S$ state normalized for $N$=1-5 Rydberg atoms, which are detected by the selective field ionization technique (Fig. 5c).

The amplitude and width of the resonance increase when the number of atoms grows, in agreement with the developed theoretical model for a disordered ensemble of interacting Rydberg atoms [19]. The appearance of resonance for $N$=1 indicates the finite detection efficiency of Rydberg atoms by the selective field ionization, as discussed in our paper [14]. The unknown detection efficiency was measured from the relationship of the amplitudes of Förster resonances for $N$=2 и $N$=1, and found to be 65%. This is a record high value for the detection of single Rydberg atoms. In our ensuing paper [20], we have also notices the effect of cold photoions Rb$^+$ on the shape of the Förster resonance. The ions appeared upon excitation of Rydberg atoms by intense broadband pulsed laser radiation. The intense radiation leads not only to the excitation of Rydberg atoms, but also to their multiphoton ionization, as was also noticed in the other paper [30]. In our study [20], the electric field of the photoions caused asymmetry, broadening and decrease in the amplitude of the Förster resonance. In order to eliminate this effect we applied a pulse of the extracting electric field at the moment of laser excitation, which resulted in a good resonance shape shown in fig. 5c.

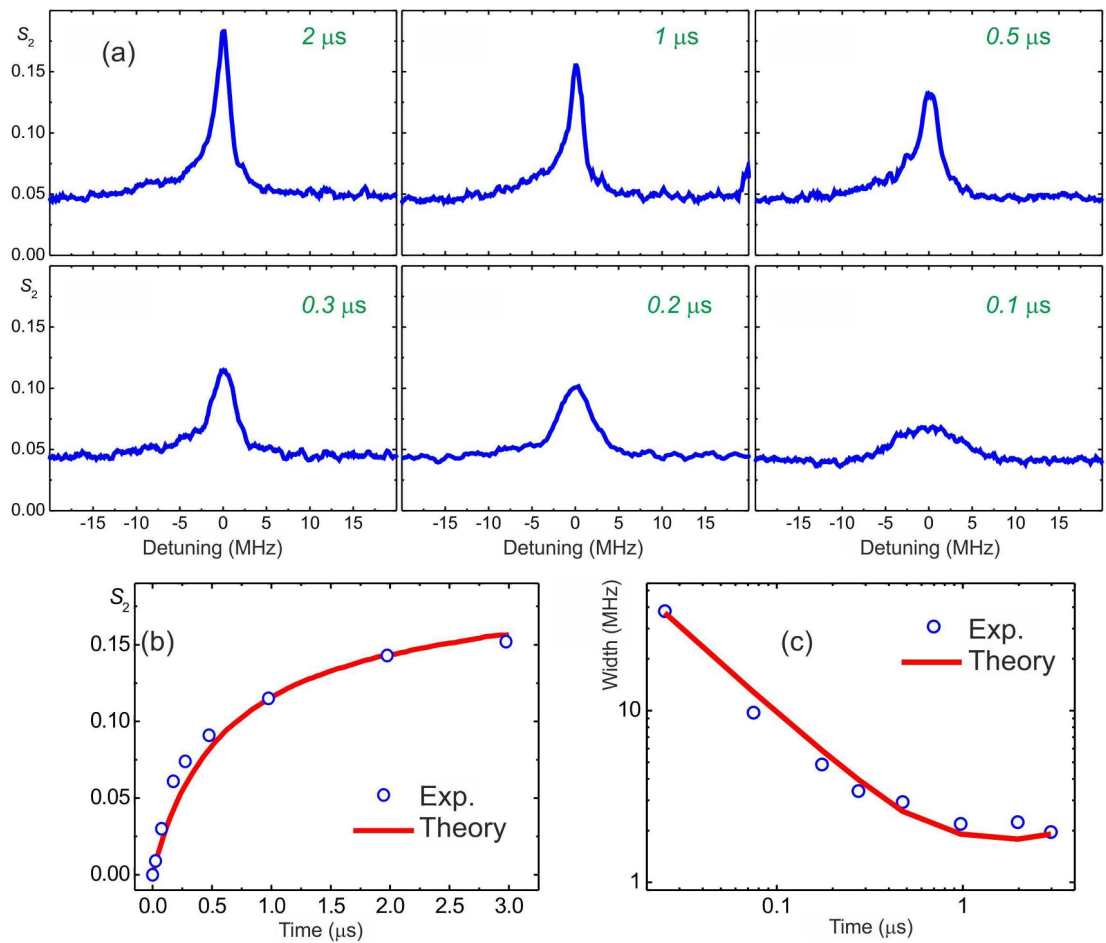

**Fig. 6.** (a) Line shape of the Förster resonance Rb(37$P$)+Rb(37$P$)→Rb(37$S$)+Rb(38$S$) for two cold Rb Rydberg atoms in the 37$P_{3/2}$(|$M_J$|=1/2) state at the various interaction times (0.1-2 µs). (b) Comparison of the experimental (circles) and theoretical (solid line) dependences of the amplitude of two-atom Förster resonance on the interaction time. (c) Comparison of the experimental (circles) and theoretical (solid line) dependences of the width of two-atom Förster resonance on the interaction time.

At present, we are carrying further studies of Förster resonances, in particular, studying their time dynamics. Figure 6a presents the line shape of the Förster resonance Rb(37$P$)+Rb(37$P$)→Rb(37$S$)+Rb(38$S$) for two cold Rb Rydberg atoms in the 37$P_{3/2}$(|$M_J$|=1/2) state at the various interaction times (0.1-2 µs). Decreasing the interaction time first leads only to a decrease in the amplitude, but at times below 1 µs it also leads to resonance broadening (Fourier width of the interaction pulse). A comparison of the experimental and theoretical results for the dependences of the amplitude and width of the two-atom Förster resonance on the interaction time is presented in Fig. 6b,c. In the theoretical calculations, we applied the same method as in paper [19], but it was modified to account for the additional parasitic broadenings of the Förster resonance using a density-matrix model with a phase relaxation as we did earlier in paper [21] for the spectra of three-photon laser excitation. Some differences between the theoretical and experimental dependences (Fig. 6b,c) point to the need for further development of the theoretical model.

Unfortunately, the Stark tuning of the Förster resonances is applicable only to a limited set of Rydberg states. For example, a narrow Stark-tuned Förster resonance in Rb atoms between neighboring Rydberg states can be obtained only for $nP_{3/2}$ states with $n \leq 38$, $nD_{3/2}$ states with $n \geq 40$, and $nD_{5/2}$ states with $n \geq 43$.

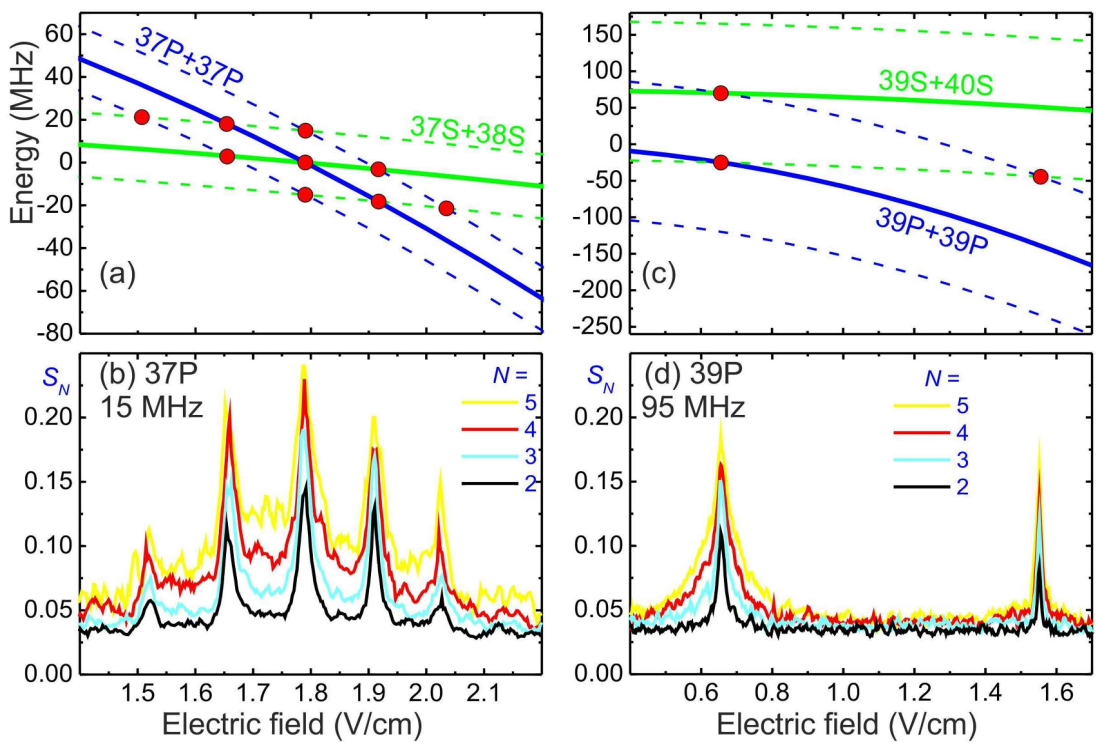

**Fig. 7.** (Color online) (a) Energy levels of the initial $37P+37P$ and final $37S+38S$ collective states of the two Rb Rydberg atoms in the electric field (solid curves) in the presence of the first additional Floquet levels detuned by $\pm 15$ МГц (dashed lines). The circles indicate the intersections of the Floquet states corresponding to the rf-induced Förster resonances. (b) Experimental record of the rf-induced Förster resonances for $N$=2-5 detected Rydberg atoms at the rf-amplitude of 100 mV. The positions of the resonances agree with the intersections of the Floquet states. (c,d) The same as in figures a and b, respectively, for the case of "inaccessible" Förster resonance for the $39P$ state in an rf-field with a frequency of 95 MHz and amplitude of 100 mV. Only the first- and second-order resonances are observed in this case.

In our paper [23], we have demonstrated that, to obtain "inaccessible" Förster resonances, which cannot be tuned by the dc electric field, one can utilize a radio-frequency (rf) electric field, and that application of the rf-field leads to effective passage from the van-der-Waals to the resonant dipole-dipole interaction (Fig. 7).

To switch the interaction on and off, we took advantage of the Stark-switching technique [31]. Laser excitation of Rydberg atoms occurs during 2 μs at the fixed value of the electric field 5.6 V cm$^{-1}$. Then the field decreases to a lower value, which is close to the resonance (1.79 V cm$^{-1}$ for the $37P_{3/2}$ state) and acts for 3 μs, following which the field increases back to 5.6 V cm$^{-1}$. The lower value of the electric field is scanned across the Förster resonance, while the selective field ionization signal accumulates for $10^3$-$10^4$ laser pulses. A radio-frequency field with the variable amplitude (0-0.3 V) and frequency (10-100 MHz) is added to the lower value of the electric field.

For the Förster resonance $\mathrm{Rb}(37P_{3/2}) + \mathrm{Rb}(37P_{3/2}) \rightarrow \mathrm{Rb}(37S_{1/2}) + \mathrm{Rb}(38S_{1/2})$, which is already familiar to us and whose collective energy levels are shown in Fig. 7a, the application of an rf-field with a frequency of 15 MHz and amplitude of 100 mV causes the appearance of several Förster resonances instead of a single one (Fig. 7b). These resonances correspond to single-photon and multiphoton rf-transitions between many-body collective states of a Rydberg quasimolecule formed by the interacting atoms. When the number of atoms $N$ rises, the resonance amplitude also grows due to the increase in the total energy of dipole-dipole interaction. Observed additional resonances have the amplitudes and widths, which are comparable to those of the central (main) resonance, if the amplitude of the rf-field is large enough. This means that the van-der-Waals interaction transforms to dipole-dipole interaction with high efficiency reaching 50-100%. By varying the frequency of the rf-field, we could control the positions of the additional peaks.

Now let us turn to the "inaccessible" Förster resonances, which cannot be tuned by a dc electric field alone. As an example, we consider the Förster resonance $\mathrm{Rb}(39P_{3/2}) + \mathrm{Rb}(39P_{3/2}) \to \mathrm{Rb}(39S_{1/2}) + \mathrm{Rb}(40S_{1/2})$ whose collective energy levels are shown in Fig. 7c. The dc electric field itself only increases the energy detuning and makes the atom interactions weaker. However, our experience with the Förster resonance for the $37P_{3/2}$ state suggests that the rf-field can induce transitions between collective states, so that Förster resonance should appear irrespective of the possibility of tuning it by the dc electric field. The dc field, however, should be applied to increase the efficiency of the rf-resonances due to the quadraticity of the Stark effect in the $S$ and $P$ states [32].

Figure 7d presents the experimental record of the Förster resonance $\mathrm{Rb}(39P_{3/2}) + \mathrm{Rb}(39P_{3/2}) \to \mathrm{Rb}(39S_{1/2}) + \mathrm{Rb}(40S_{1/2})$ in the rf-field with a frequency of 95 MHz and amplitude of 100 mV. Increasing the frequency shifts the resonance to a higher dc electric field. The resonance efficiency is rather high because its amplitude is comparable to that of the "ordinary" resonance, and at $N$=5 it is close to the maximum value 0.25 for a disordered atom ensemble [19]. Figure 7 thus evidences the possibility of transferring the van-der-Waals interaction to the dipole-dipole one with high efficiency using a dc electric field and the rf-field added.

Let us now turn to the physical interpretation of the rf-induced Förster resonances. On the one hand, the rf-field induces direct transitions between collective states of the interacting Rydberg atoms. A few rf-photons of frequency $\omega$, being absorbed or emitted by a quasimolecule consisting of a few Rydberg atoms, can compensate for the Förster energy defect $\Delta$ if its value is a multiple of $\omega$. On the other hand, the rf-induced Förster resonances can also be explained in terms of additional Floquet levels appearing upon periodic perturbation of the Rydberg states by the rf-field due to the Stark effect [32]. The Floquet levels comprise an infinite set of additional levels detuned from the original level by the frequency, which is multiple of the modulation frequency $\omega$, while their amplitudes are described by the generalized Bessel functions. In this interpretation, the rf-induced Förster resonances occur not for the ordinary Rydberg levels, but for the energy-detuned Floquet levels, which satisfy the resonance condition $\Delta = m\omega$, where $m$ is an integer. They undergo crossings at some certain values of the dc electric field.

Figure 7a shows the energy levels of the initial $37P+37P$ and final $37S+38S$ collective states of the two rubidium Rydberg atoms in the electric field (solid curves) in the presence of the first Floquet levels detuned by $\pm15$ MHz (dashed curves). The circles indicate the intersections of the Floquet levels corresponding to the Förster resonances induced by the rf-field. These resonances are well seen in Fig. 7b. Similar resonances are shown in Fig. 7c,d for the state $39P_{3/2}$ in the rf-field with a frequency of 95 MHz and amplitude of 100 mV. There are many fewer intersections of the Floquet levels for these "inaccessible" Förster resonances. A narrow second-order resonance at 1.55 V cm$^{-1}$ is clearly seen along with the much stronger first-order resonance at the electric field 0.66 V cm$^{-1}$. The first-order resonance saturates and broadens as $N$ grows, while there is no saturation for the second-order resonance and it remains narrow for all $N$. We have also observed for the first time additional Rydberg Floquet levels in the spectra of three-photon excitation of the $37P$ state in the rf-field with a frequency of 15 MHz [23].

Radio-frequency assisted Förster resonances and Floquet levels can be especially useful to enhance the dipole blockade effect in mesoscopic Rydberg ensembles, which is used to perform two-qubit gates and in a number of other applications [27]. For $nP_{3/2}$ states with $n$=40-100 the required radio-frequencies lie in the 100-325 MHz range, and for $nS$ states with $n$=70-120 they are in the 140-700 MHz range. These are relatively low frequencies, which are convenient for making use in experiments.

The rf-induced Förster resonances thus significantly extend the set of Rydberg states suitable for implementing long-range resonant dipole-dipole interaction. In fact, by using an rf-field one can enhance the interactions of almost arbitrary Rydberg atoms with large principal quantum numbers.

## 4. Application of cold Rydberg atoms to quantum information processing.

Experimental realization of quantum computing is one of the most interesting issues in the modern physics, mathematics and informatics. Quantum information was established in 1973, when Soviet mathematician Alexander Holevo has published an estimate of the amount of information which could be transferred through a quantum communication channel. This work is known now as Holevo's theorem [33]. In 1976 polish mathematician Roman Stanislav Ingarden published a paper "Quantum theory of information" [34]. In 1980 Yury Manin in his book "Computable and Incomputable" discussed the using of "quantum automata" to simulate complex processes like DNA replication [35].

The idea of creating a quantum computer became popular after the Nobel Prize laureate Richard Feynman addressed this subject. In 1981 during the first conference on computational physics at the Massachusetts Institute of Technologies he delivered a lecture "Simulation of physics with computers" where he posed a model of a quantum computer. The next important step was done by P. Shore only in 1994 when the quantum algorithm of factorization of large numbers was developed [36]. This algorithm is the most attractive for practical implementation of quantum computing because it can be used to decrypt secret messages. In the same year, an experimental quantum computation system with ultracold ions was proposed [37]. The two-qubit "Controlled-NOT" (CNOT) quantum gate was successfully demonstrated experimentally in the next year [38]. In 1998 Grover's quantum algorithm for searching in an unsorted database [39] was implemented on a two-qubit quantum computer using nuclear magnetic resonance [40].

Later on, a number of the physical systems [41-46] were considered for experimental implementation of quantum computing, including Josephson junctions in superconductors [47], photons [48], quantum dots [49] and atoms in optical lattices and optical dipole traps [11,50]. The applicability of each physical realization to quantum computing can be estimated using the five criteria, proposed by David DiVincenzo [43,44] as follows:

1. A quantum register must comprise many qubits represented by some quantum system. Individual qubits must be distinguishable, the quantum state of each qubit must be externally controlled (individual addressing), and each qubit must be a two-level system which does not decay spontaneously to any third level. It should also be possible to add more qubits to the quantum register if necessary (scalability).
2. The quantum register must be initialized prior to the beginning of quantum computation. For example, all qubit must be initially prepared in quantum state $|0\rangle$.
3. The decoherence time must be sufficiently large, at least $10^4$ longer than the duration of one quantum gate.
4. Single-qubit and two-qubit quantum gates must be reversible and described by the unitary matrices to be performed over single qubits. Examples are single-qubit rotation of the quantum state of a qubit on a Bloch sphere by an arbitrary angle and the two-qubit CNOT gate, which creates an entangled state of the two qubits.
5. The final state of the quantum register must be measured quickly and with high quantum efficiency.

Ultracold neutral atoms can be utilized to implement quantum gates, since they perfectly meet all of the DiVincenzo criteria. The two qubit states can be the long-lived hyperfine sublevels of the ground state of ultracold alkali-metal atoms, which are preliminarily laser cooled and then captured in dipole traps.

Experimental realization of single-qubit quantum gates is not a big difficulty for many physical systems. The qubit wave function can be described as a superposition of logical states $|\psi\rangle = a|0\rangle + b|1\rangle$. The transformation of the wave function following from quantum gates is described by matrices 2×2 in dimension. Examples of such gates are the Hadamard gate H and the phase gate $U(\varphi)$:

$$H = \frac{1}{\sqrt{2}}\begin{pmatrix} 1 & 1 \\ 1 & -1 \end{pmatrix}, \quad U(\varphi) = \begin{pmatrix} 1 & 0 \\ 0 & e^{i\varphi} \end{pmatrix}, \tag{1}$$

In order to implement two-qubit quantum gates, one needs to control the interaction of two qubits. For neutral atoms this can be achieved by their temporary excitation to Rydberg states. In 2001, M.D.Lukin et al. [27] have considered for the first time the effect of a dipole blockade. Only one atom in an ensemble of the interacting atoms can be excited to a Rydberg state by narrow-band laser radiation. This effect results from the shift of collective energy levels in the atomic system due to interatomic interaction. The collective states of the atomic ensemble, containing several Rydberg excitations, are shifted out of resonance with laser radiation. The effect of dipole blockade can be used to store quantum information in the collective states of atomic ensembles.

Another application is implementation of two-qubit quantum gates. Two-qubit quantum gates are the most important for a universal quantum computer. In the basis of two-qubit states $|00\rangle$, $|01\rangle$, $|10\rangle$, $|11\rangle$ the quantum gates are described by matrices 4×4 in dimension. The CNOT gate inverts the state of the target qubit if the control qubit is prepared in state $|1\rangle$. The controlled-phase (CZ) gate alters the phase of the target qubit if the control qubit is prepared in state $|1\rangle$. The CNOT and CZ gates are described in the basis $|00\rangle$, $|01\rangle$, $|10\rangle$, $|11\rangle$ by the following matrices:

$$CNOT = \begin{pmatrix} 1 & 0 & 0 & 0 \\ 0 & 1 & 0 & 0 \\ 0 & 0 & 0 & 1 \\ 0 & 0 & 1 & 0 \end{pmatrix} \qquad CZ = \begin{pmatrix} 1 & 0 & 0 & 0 \\ 0 & 1 & 0 & 0 \\ 0 & 0 & 1 & 0 \\ 0 & 0 & 0 & -1 \end{pmatrix}. \tag{2}$$

Substantial progress in the investigations of Rydberg atoms to implement quantum computation has been made by the team of Mark Saffman (University Wisconsin-Madison, USA). In 2010 this team demonstrated for the first time two-qubit quantum gates using a dipole blockade [51]. A diagram of their experiment is shown in Fig. 8a. Two Rb atoms were trapped in two optical dipole traps formed by tightly focused laser beams from a neodymium laser (1064 nm wavelength) with a waist radius of 3 μm. The distance between the dipole traps measured 10 μm. The hyperfine sublevels $|0\rangle = |5S_{1/2}, F = 1, M_F = 0\rangle$ and $|1\rangle = |5S_{1/2}, F = 2, M_F = 0\rangle$ of the ground state of the $^{87}$Rb atom were used as the logical states of a qubit. Optical pumping to state $|1\rangle$ was used to initialize the qubits. Single-qubit gates were implemented utilizing Raman transitions between the qubit states with a detuning of 40 GHz from the intermediate excited state $5P_{3/2}$.

To implement two-qubit gates the atoms were excited to the Rydberg state $97D_{5/2}$ by the radiation of two lasers with wavelengths of 780 nm and 480 nm. A diagram for implementing a CNOT gate is illustrated in Fig. 8b. A sequence of laser pulses inverted the state of the target qubit, but only when the control qubit was in state $|1\rangle$ and could not be excited to the Rydberg state. Otherwise the sequence of inverting pulses had no effect on the target qubit, because laser excitation of Rydberg atoms was impossible due to a dipole blockade.

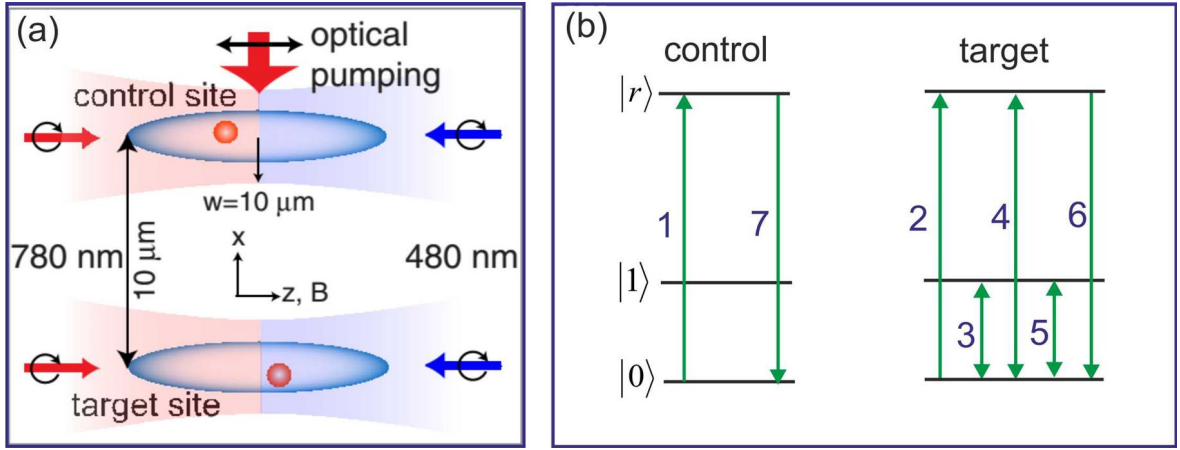

**Fig. 8.** (a) Schematic of the experiment by M.Saffman's team [51] on the implementation of a two-qubit CNOT gate. Two Rb atoms are captured in two optical dipole traps, which are located 10 μm apart from each other. Laser radiation with wavelengths 780 nm and 480 nm is used for coherent two-photon excitation of the Rydberg states. Only one Rydberg atom can be excited due to a dipole blockade. (b) Diagram of implementation of the CNOT gate. The sequence of pulses 2-6 inverts the state of the target qubit, if control qubit is initially in state $|1\rangle$ and is not therefore excited to the Rydberg state. Pulses 1 and 7 excite the control qubit to the Rydberg state if it is initially in state $|0\rangle$. In this case pulses 2-6 do not invert the target qubit.

Simultaneously with the Saffman's team, the entangled states of two atoms were demonstrated by A.Browaeys' group using the same dipole blockade effect [52]. In their study, Rb atoms were trapped in optical dipole traps located 4 μm form each other. The atoms were initially prepared in a hyperfine sublevel of the ground state $|1\rangle = |5S_{1/2}, F = 2, M_F = 0\rangle$. For two-photon Rydberg excitation $|1\rangle = |5S_{1/2}, F = 2\rangle \rightarrow |5P_{1/2}, F = 2\rangle \rightarrow |r\rangle = |58D_{3/2}, F = 3\rangle$ the two lasers with wavelengths of 795 nm and 475 nm were employed. Due to the dipole blockade, a collective state $\frac{1}{\sqrt{2}}\left(|1,r\rangle + |r,1\rangle\right)$ with only a single Rydberg atom was excited. To create an entangled state of two atoms originally populating different hyperfine sublevels of the ground-state, an additional laser pulse was used on the transition $|r\rangle \rightarrow |0\rangle = |5S_{1/2}, F = 1\rangle$. Finally, the entangled quantum Bell state $|\Psi\rangle^+ = \frac{1}{\sqrt{2}}\left(|1,0\rangle + |0,1\rangle\right)$ was obtained. Entangled states using the dipole blockade were also demonstrated in experiments by M.Saffman's team [53].

The next step for Saffman's team was concerned with building a quantum register based on 49 optical dipole traps with single Cs atoms in each trap, as shown in Fig. 9a [54,55]. In contrast to the previous work by this team, where the atoms were trapped in the maxima of laser intensity, the atoms in Refs. [54,55] were trapped in the local minima of a laser field intensity. For this purpose, a laser radiation with a wavelength of 780 nm was used, which was blue-detuned from the resonance in Cs atoms. In addition, a special configuration of those laser beams with different frequencies and polarizations was prepared to avoid formation of an undesirable interference pattern. An array of optical dipole traps was created invoking active diffraction elements to control the wave front of the laser beams. The distance between adjacent traps was 3.8 μm in an array with the 7×7 dimensions. This ensures a high density of the qubits. A quantum register with $10^6$ qubits would occupy an area less than 1 square centimeter, as noted by the authors.

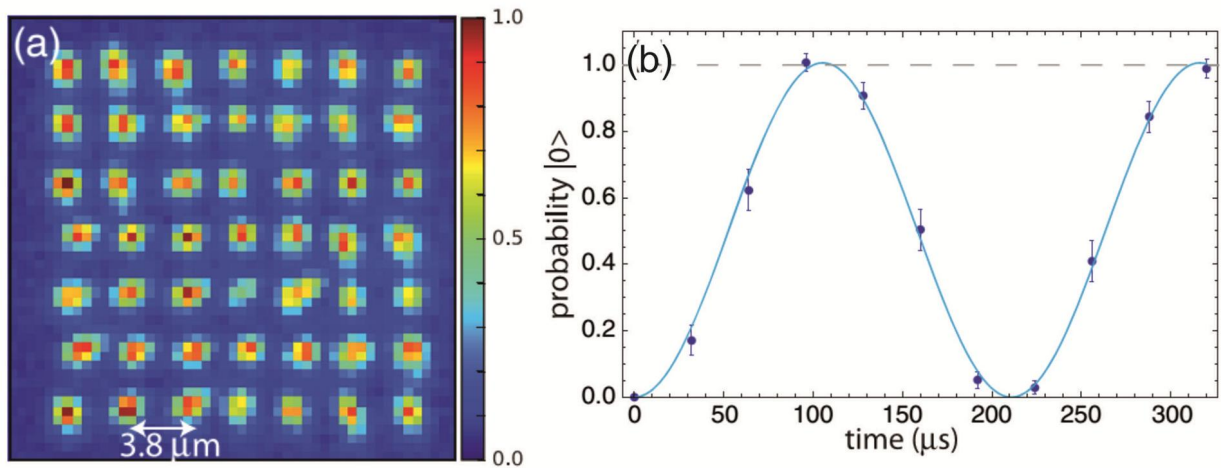

**Fig. 9.** (a) Observed fluorescence of an array of optical dipole traps loaded with single Cs atoms, used by M.Saffman's team to build a quantum register [54,55]. The image is obtained by averaging over 500 measurements. (b) Single-qubit Rabi oscillations observed when the resonant microwave radiation interacts with hyperfine sublevels of the ground state of single Cs atoms.

In this quantum register, single-qubit rotation was realized employing microwave radiation with a frequency of 9.19 GHz (a magneto-dipole transition between the states $|0\rangle = |6S_{1/2}, F = 3, M_F = 0\rangle$ and $|1\rangle = |6S_{1/2}, F = 4, M_F = 0\rangle$). The fidelity of single-qubit rotation was higher than 99% [55]. Rabi oscillations in the population of state $|0\rangle$ were observed in the experiment (Fig. 9b). Individual addressing to separate atoms by microwave radiation is possible if the energy levels of the atom are shifted by tightly focused nonresonant laser radiation due to the ac Stark effect, and in this way the atom is put into a resonance with the microwave field, which is originally detuned from the microwave transition [56]. The addressing of the individual qubits by laser radiation is achieved by using acousto-optical modulators which scan the beam direction.

Experimental realization of two-qubit gates which create entangled states of two atoms is much more difficult. In recent experiments by M.Saffman's team [57] the fidelity of the Bell states was still less than 80%. One of the error sources is the light shift, which appears during two-photon excitation of Rydberg atoms. The loss of single atoms in the optical traps also plays an important role in the reduction of fidelity. This loss may be partially compensated for by post-selection, when only those events in which no loss occurs are sampled.

The results of recent experiments show that high fidelity of two-qubit quantum gates with neutral cold atoms requires precise control and optimization of the parameters of the experiment and a deeper understanding of the physics of laser excitation and interaction of multilevel atoms in intense laser fields. Potential systems for realizing two-qubit quantum gates with neutral atoms can be based not only on the dipole blockade, but also on a phase shift which results from the dipole-dipole interaction of two Rydberg atoms. For example, a scheme of two-qubit controlled phase gate was proposed by us in Ref. [28]. In this scheme, dipole-dipole interaction leads to coherent oscillations of populations of collective states of two interacting Rydberg atoms initially excited to states of opposite parity (*nS* and *nP*). This results in a phase shift of the collective wave function during resonant dipole-dipole interaction.

Coherent oscillations of the collective states of two interacting Rydberg atoms in the conditions of Förster resonance can be used similarly. These oscillations have already been observed in the experiment by A.Browaeys's group [58]. In the last work, two $^{87}$Rb atoms were trapped in two optical dipole traps. The Förster resonance was observed for the states $\left| p = 61P_{3/2}, m_j = 1/2 \right\rangle$, $\left| d = 59D_{3/2}, m_j = 3/2 \right\rangle$ and $\left| f = 57F_{5/2}, m_j = 5/2 \right\rangle$ (Fig. 10). Exact Förster resonance, when the collective states $\left| pp \right\rangle$ and $\left| df \right\rangle$ became degenerate, appeared in a weak electric field of 32 mV/cm. Coherent oscillations of the population of the collective states of two Rydberg atoms were observed experimentally. The frequency of the oscillations was determined by the interaction energy. This was confirmed by a reduction in the frequency with an increase in the distance between the optical traps.

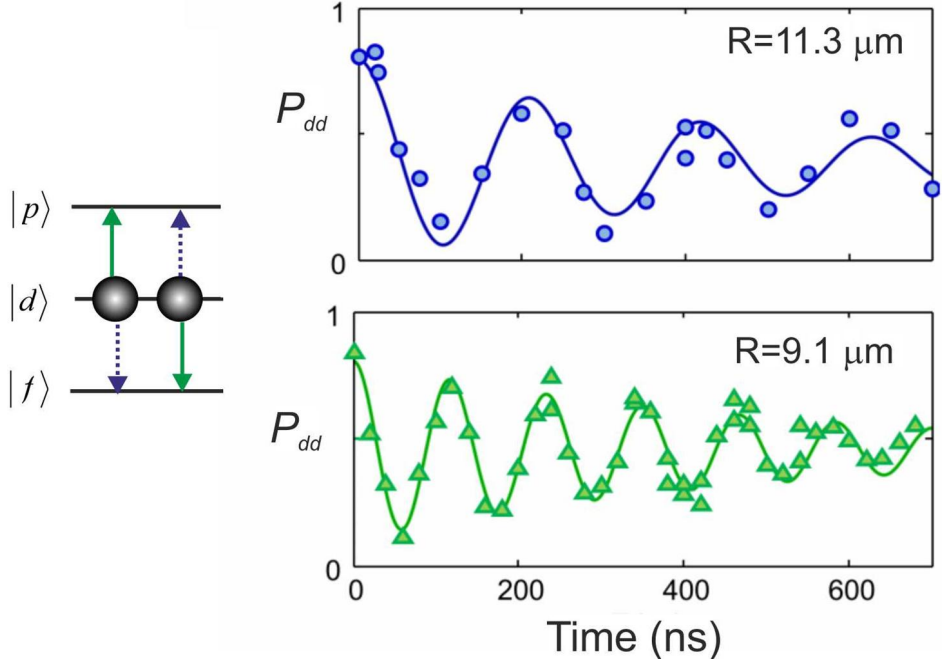

**Fig. 10**. Collective oscillations of populations in a system of two interacting atoms, induced by resonant dipole-dipole interaction [58].

In general, the Stark-tuned Förster resonances have been known for long time. They were involved, for instance, for the first reliable observation of the dipole blockade [59]. At present, they remain the subject of intense theoretical and experimental research because of their importance for applications in quantum information [22,59-65] due to the ability to efficiently controlling the strength of Rydberg interaction in a weak electric field.

## 5. Schemes of the quantum gates and quantum simulations based on the dipole blockade effect in mesoscopic atomic ensembles.

Based on the dipole blockade effect [27], we have proposed a new method of deterministic single-atom Rydberg excitation in the sites of optical lattices initially loaded with a random number of atoms [18]. This method can be applied for single-atom loading of the sites of optical lattices and building quantum registers with neutral atoms. We have found that Rydberg excitation in mesoscopic atomic ensembles with chirped laser pulses (with rapidly changed frequency) in the regime of a dipole blockade occurs adiabatically and is insensitive to variations of the collective Rabi frequency of single-atom excitation. The frequency of Rabi oscillations in the dipole blockade regime depends on the unknown number $N$ of atoms in the ensemble as $\Omega_N = \Omega_1 \sqrt{N}$ (here $\Omega_1$ is the Rabi frequency for a single atom). Our numeric simulations have shown that if interatomic interaction is strong enough to ensure the full blockade regime, the illumination of the $N$-atom ensemble by a chirped laser pulse leads to the deterministic excitation of only one atom with a probability close to 1, and proceeds as practically independent of $N$ [18].

A similar effect can be observed if we make use of two-photon excitation with Stimulated Rapid Adiabatic Passage (STIRAP), which is in many aspects equivalent to chirped excitation. The STIRAP method is based on a counter-intuitive sequence of laser pulses, when the laser pulse on the second excitation step arrives earlier, than the pulse at the first step [66]. This leads to adiabatic excitation due to time-dependent dynamic shifts of dressed energy levels of the two-photon transition.

Based on this method, we have developed the original schemes of quantum gates for mesoscopic atomic ensembles [21,23] which can be used, in particular, to implement a measurement-based quantum computing (MBQC) [67]. We propose using a two-dimensional array of optical dipole traps, each loaded with a random number of atoms instead of single atoms, as a quantum register. A collective qubit is represented by an $N$-atom ensemble with the logical states

$$\left|\overline{0}\right\rangle = \left|000...000\right\rangle \text{ and } \left|\overline{1}\right\rangle = \frac{1}{\sqrt{N}} \sum\nolimits_{j=1}^{N} \left|000...1_j...000\right\rangle,$$

where states $\left|0\right\rangle$ and $\left|1\right\rangle$ stand for hyperfine sublevels of the ground state in a single atom. Collective states of the ensemble are bound through the deterministic single-atom Rydberg excitation of state

$$\left|\overline{r}\right\rangle = \frac{1}{\sqrt{N}} \sum\nolimits_{j=1}^{N} \left|000...r_j...000\right\rangle,$$

which contains only one Rydberg atom in the ensemble.

As dipole blockade enables only one Rydberg excitation, states $\left|\overline{0}\right\rangle$ and $\left|\overline{r}\right\rangle$ are bound through an optical transition with the collective Rabi frequency $\Omega_N = \Omega_1 \sqrt{N}$. At the same time, states $\left|\overline{1}\right\rangle$ and $\left|\overline{r}\right\rangle$ are bound through an optical transition with single-atom Rabi frequency $\Omega_1$. It is also possible to use additional Rydberg states coupled by microwave transitions, or to make use of additional hyperfine sublevels of the ground state in order to limit the scheme by optical transitions from the ground to Rydberg states only (Fig. 11).

As a result, single-qubit and two-qubit quantum gates with mesoscopic qubits can be realized as a sequence of laser and microwave pulses shown in Fig. 11. These schemes also employ a method of dynamic phase compensation developed by us in studies [22,24]. The dynamic phase is accumulated during adiabatic passage and it generally depends on $N$. In order to compensate for it, we address a combination of the forward and reverse pulse sequences, which resembles a photon echo schemes.

The schemes of quantum gates performed with an auxiliary microwave transition between Rydberg states are shown in Fig. 11a-c. The schemes addressing all-optical transition gates are shown in Fig. 11d-f. A detailed description of these schemes can be found in our papers [22,24]. In particular, we analyze there the fidelity of quantum gates with mesoscopic qubits. Our calculations demonstrated that in experiments with a two-photon Rydberg excitation of Rb and Cs atoms using STIRAP, the detuning from the intermediate excited state must be larger than 2 GHz to achieve an error below 0.5%.

We now consider an example of quantum simulations with mesoscopic atomic ensembles in the dipole blockade regime proposed by us in paper [25]. It is well known that the fluctuations of the photon number in the quantized electromagnetic field mode lead to nonclassical features in the time dynamics of populations of a two-level atom interacting with this field, namely, to collapses and revivals of Rabi oscillations (Jaynes-Cummings model) [68]. In this model, the frequency of

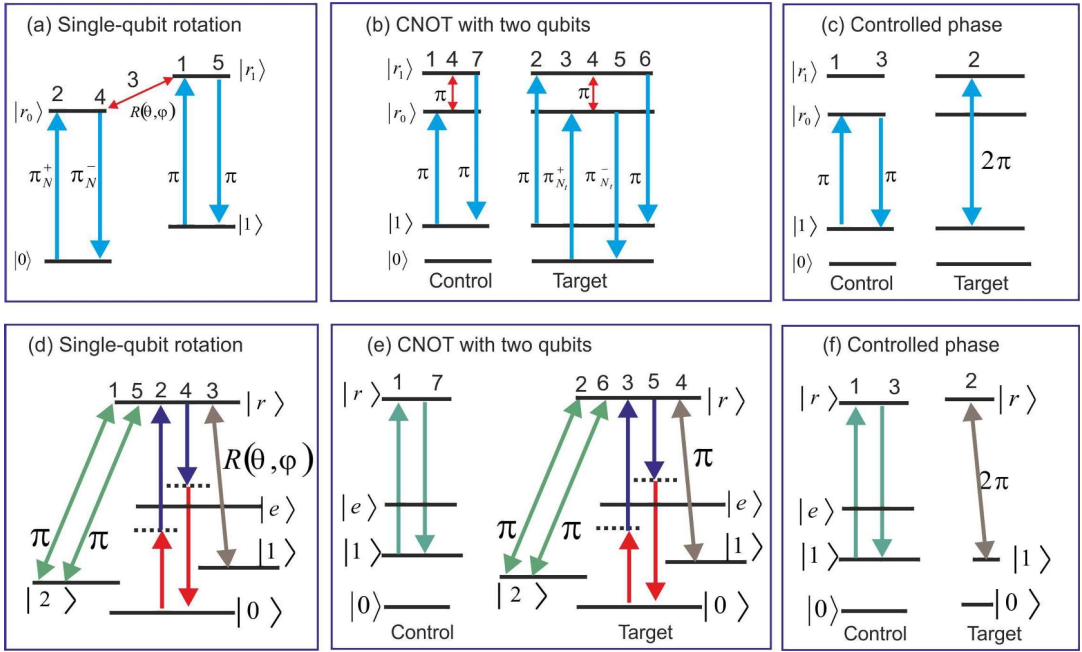

**Fig. 11.** Schemes of the quantum gates with mesoscopic atomic qubits comprised an unknown random number of atoms in the ground state. Using an auxiliary microwave transition between Rydberg states, we can perform: (a) single-qubit rotation on an arbitrary angle, (b) a two-qubit "Controlled NOT" (CNOT) gate, and (c) a two-qubit "Controlled phase" (CZ) gate. Using only optical transition, we can perform: (d) single-qubit rotation on an arbitrary angle, (e) a two-qubit "Controlled NOT" (CNOT) gate, and (f) a two-qubit "Controlled phase" (CZ) gate.

Rabi oscillations between the states of a two-level quantum system is found proportional to $\sqrt{n}$, where $n$ is an unknown and random number of photons in the mode of a quantized electromagnetic field which is resonant to the atomic transition (Fig. 12a). For a coherent state of the electromagnetic field with a random (Poissonian) distribution of the photon number, the Jaynes-Cummings model demonstrates dephasing and subsequent rephasing of the Rabi oscillations due to quantum interference of the Rabi frequencies corresponding to different numbers of photons [13].

We suggested that similar dynamics can be ascribed to the probability of single-atom excitation in mesoscopic atomic ensembles with a random number of atoms $N$, which interact with the resonant laser radiation in the Rydberg dipole blockade regime. The mesoscopic ensembles in the regime of a perfect blockade can be described as an effective two-level system whose two levels are the collective Dicke states $|G\rangle$ and $|R\rangle$ (Fig. 12b). The collapses and revivals of collective oscillations of populations of the Dicke states appear due to the $\sqrt{N}$ dependence of the collective Rabi frequency of single-atom Rydberg excitation in the Rydberg blockade regime, which is fully analogous to the Jaynes-Cummings model.

The results of our simulations of the time dependence of probability $P_i$ of exciting $i$ Rydberg atoms in a mesoscopic ensemble with the mean number of atoms $\overline{N}$ =7 are shown in Figs. 12c-f. The atoms are supposed to be randomly distributed in an optical dipole trap with the radius $r$=2-5 μm. Model simulations have been performed for Cs(80S) atoms with the van-der-Waals interaction constant $C_6/(2\pi) = 3.2 \times 10^6$ MHz·μm$^6$. The effective dipole blockade radius can be found by equating the van-der-Waals interaction energy to the Rabi frequency: $R = \left(C_6/\Omega\sqrt{N}\right)^{1/6} = 10$ μm for $N$=7 atoms. Depending on the size of the optical dipole trap, the dipole blockade for such atoms can be complete or partial.

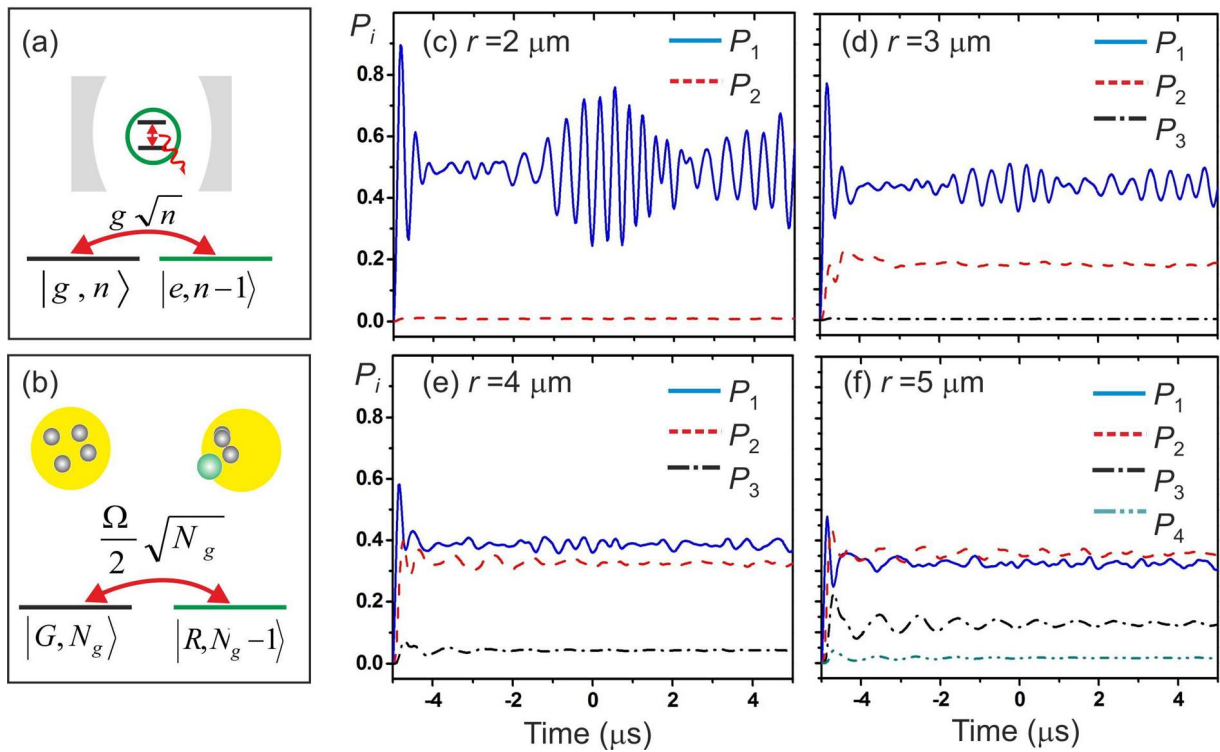

**Fig. 12.** (a) Interaction of a single atom with a quantized electromagnetic field in a cavity is described by the Jaynes-Cummings model [68]. There are two coupled states of the atom+field system, while the number of photons $n$ is random. (b) Schemфeшc of the coupled states of ф mesoscopic $N$-atom ensemble in the Rydberg blockade regime, which is equivalent to the Jaynes-Cummings model. (c-f) Numerically simulated time dependences of the probability $P_i$ of exciting $i$ Cs(80S) Rydberg atoms in a mesoscopic ensemble with $\bar{N}$ =7 atoms, randomly distributed in an optical dipole trap with radius (c) $r$=2 µm, (d) $r$=3 µm, (e) $r$=4 µm, and (f) $r$=5 µm.

In Fig 12c the collapses and revivals in the probability $P_1$ are clearly seen at $r = 2$ µm for a time interval shorter than 10 µs (solid curve). The probability $P_2$ of exciting two Rydberg atoms (dashed curve in fig. 12c) is close to zero, which means that the dipole blockade is complete. The time dependence of $P_1$ perfectly matches the Jaynes-Cummings dynamics. However, at $r = 3$ µm the calculated revivals are substantially weaker due to smaller interaction energy of Rydberg atoms and the incompleteness of the dipole blockade, as seen in Fig. 12d (solid curve), despite the fact that the trap size is still less than the blockade radius. The breakdown of a complete blockade also appears in this figure as an increase in the probability $P_2$ of exciting two Rydberg atoms (dashed curve). The results of calculations for the larger sizes of the trap are shown in Fig. 12e-f, which demonstrates that the revivals have vanished. The interaction between the atoms in the ensemble is too weak. As a result, we should observe independent single-atom Rabi oscillations, but they are washed out due to averaging over time and frequency.

Besides the quantum simulation of the Jaynes-Cummings dynamics, experimental observation of collapses and revivals can be used for identification of the regime of the perfect dipole blockade without the need to measure the actual number of detected Rydberg atoms. This is of interest to quantum information with mesoscopic ensembles comprising a random number of atoms.

## 6. Conclusions.

Usage of ultracold trapped atoms opens up the new possibilities for various basic and applied researches in the field of atomic spectroscopy due to spatial localization of atoms in a small volume and the absence of the transit-time and Doppler broadenings. Cold atoms in optical traps are recognized as promising objects to create qubits of a quantum computer. One of the new directions in atomic spectroscopy is the experimental study of ultracold atoms in Rydberg states. Due to their unique properties it is possible to conduct experiments with single or a few atoms. This is extremely important for applications in quantum information processing. For example, based on the strong long-range interactions of Rydberg atoms having large dipole moments, it is possible to implement two-qubit quantum gates, realize single-atom loading of optical traps, and implement schemes of quantum simulations in the ensembles of interacting particles. The most active studies in this area are being conducted abroad by the research teams of M.Saffman (USA) and A.Browaeys (France).

We have performed a number of experimental and theoretical studies on the spectroscopy of cold Rb Rydberg atoms in a small excitation volume in a magneto-optical trap. We experimentally obtained the spectra of the three-photon excitation of Rydberg $nP$ states, studied the interactions of a few Rydberg atoms in the conditions of Förster resonance in an electric field, and implemented the enhancement of the interactions of Rydberg atoms in a radio-frequency field. We have put forward new theoretical ideas on Doppler- and recoil-free three-photon laser excitation of Rydberg atoms, the deterministic excitation of single Rydberg atoms by chirped laser pulses in the dipole blockade regime, the implementation of quantum gates over qubits represented by mesoscopic ensembles with an unknown number of atoms, and simulations of the Jaynes-Cummings dynamics in optical lattices with an unknown number of atoms. The results obtained are important for the application of cold Rydberg atoms in quantum information processing.

This study was financially supported by RFBR grants No. 13-02-00283, 14-02-00680 and 16-02-00383, programs of RAS and SB RAS, EU FP7-PEOPLE-2009-IRSES "COLIMA" (Coherent manipulation of light and matter via interferences of laser-dressed states), the Russian Quantum Center, and Novosibirsk State University.

## References

[1]  Cohen-Tannoudji C N *Rev. Mod. Phys.* **70** 707 (1998)
[2]  Metcalf H, Van Der Straten P *Laser cooling and trapping* (New York: Springer, 1999)
[3]  Grimm R, Weidemüller M, and Ovchinnikov Yu B *Adv. At. Mol. Opt. Phys.* **42** 95 (2000)
[4]  Ryabtsev I I et al. *Phys. Rev. Lett.* **104** 073003 (2010)
[5]  Chapovsky P L *JETP Lett.* **95** 132 (2012)
[6]  Goncharov A N et al. *Quantum Electronics* **44** 521 (2014)
[7]  Taichenachev A V et al. *Phys. Rev. Lett.* **97** 173601 (2006)
[8]  Sukachev D et al. *Phys. Rev. A* **82** 011405(R) (2010)
[9]  Makhalov V, Martiyanov K, Turlapov A *Phys. Rev. Lett.* **112** 045301 (2014)
[10]  Zelener B B et al. *JETP Lett.* **100** 366 (2014)
[11]  Brennen G K et al. *Phys. Rev. Lett.* **82** 1060 (1999)
[12]  Gallagher T F *Rydberg atoms* (Cambridge: Cambridge Univ. Press, 1994)
[13]  Beterov I M, Lerner P B Sov. Phys. Usp. **32** 1084 (1989)
[14]  Ryabtsev I I et al. *Phys. Rev. A* **76** 012722 (2007)
[15]  Viteau M et al. *J. Phys. B* **43** 155301 (2010)
[16]  Tretyakov D B et al. *JETP* **108** 374 (2009)
[17]  Ryabtsev I I et al. *Phys. Rev. A* **82** 053409 (2010)
[18]  Beterov I I et al. *Phys. Rev. A* **84**, 023413 (2011)
[19]  Ryabtsev I I et al. *Phys. Rev. A* **84** 053409 (2011)
[20]  Tretyakov D B et al. *JETP* **114** 14 (2012)
[21]  Entin V M et al. *JETP* **116** 721 (2013)
[22]  Beterov I I et al. *Phys. Rev. A* **88** 010303(R) (2013)
[23]  Tretyakov D B et al. *Phys. Rev. A*, **90**, 041403(R) (2014)

[24] Beterov I I et al. *Laser Physics* **24** 074013 (2014)

[25] Beterov I I et al. *Phys. Rev. A* **90** 043413 (2014)

[26] Jaksh D et al. *Phys. Rev. Lett.* **85** 2208 (2000)

[27] Lukin M D et al. *Phys. Rev. Lett.* **87** 037901 (2001)

[28] Ryabtsev I I, Tretyakov D B, Beterov I I *J. Phys. B* **38** S421 (2005)

[29] Saffman M, Walker T G, and Mølmer K *Rev. Mod. Phys.* **82** 2313 (2010)

[30] Viteau M et al. *J. Phys. B* **43** 155301 (2010)

[31] Ryabtsev I I, Tretyakov D B, and Beterov I I *J. Phys. B* **36** 297 (2003)

[32] van Ditzhuijzen C S E, Tauschinsky A, and van Linden van den Heuvell H B *Phys. Rev. A* **80** 063407 (2009)

[33] Holevo A S *Probl. Peredachi Inform.* **9** (3) 3 (1973) (in Russian)

[34] Ingarden R S *Rep. Math. Phys.* **10** 43 (1976)

[35] Manin Yu I *Vychislimoe i Nevychislimoe* (Calculable and Incalculable) (Moscow: Sov. Radio, 1980) (in Russian)

[36] Shor P W *SFCS '94 Proceedings of the 35th Annual Symposium on Foundations of Computer Science* [IEEE Computer Society Washington, DC, USA, 1994] p. 124

[37] Cirac I and Zoller P *Phys. Rev. Lett.* **74** 4091 (1995)

[38] Monroe C et al. *Phys. Rev. Lett.* **75** 4714 (1995)

[39] Grover L K *Proceedings, 28th Annual ACM Symposium on the Theory of Computing, Philadelphia, PA, USA, May 1996*, p. 212

[40] Jones J A, Mosca M, Hansen R H *Nature* **393**, 344 (1998)

[41] Valiev K A, Kokin A A *Kvantovye Komp'yutery: Nadezhdy i Real'nost'* (Quantum Computers: Hopes and Reality) (Izhevsk: RKhD, 2001) (in Russian)

[42] Valiev K A *Phys. Usp.* **48** 1 (2005)

[43] DiVincenzo D P *Science* **270** 255 (1995)

[44] DiVincenzo D P *Fortschr. Phys.* **48** 771 (2000)

[45] Vandersypen L M K and Chuang I L *Rev. Mod. Phys.* **76** 1037 (2004)

[46] Leibfried D et al. *Rev. Mod. Phys.* **75** 281 (2003)

[47] Makhlin Y, Schön G, and Shnirman A *Rev. Mod. Phys* **73** 357 (2001)

[48] Knill E, Laflamme R and Milburn G J *Nature* **409** 46 (2001)

[49] Loss D and DiVincenzo D P *Phys. Rev. A* **57** 120 (1998)

[50] Garcıa-Ripoll J J, Zoller P, and Cirac J I *J. Phys. B* **38** S567 (2005)

[51] Isenhower L et al. *Phys. Rev. Lett.* **104** 010503 (2010)

[52] Wilk T et al. *Phys. Rev. Lett.* **104** 010502 (2010)

[53] Zhang X L et al. *Phys. Rev. A* **82** 030306(R) (2010)

[54] Piotrowicz M J et al. *Phys. Rev. A* **88** 013420 (2013)

[55] Xia T et al. *Phys. Rev. Lett.* **114** 100503 (2015)

[56] P.Schauß et al., Science **347** 1455 (2015)

[57] Maller K M et al. *Phys. Rev. A* **92** 022336 (2015)

[58] Ravets S et al. *Nature Physics* **10** 914 (2014)

[59] Comparat D and Pillet P *J. Opt. Soc. Am. B* **27** A208 (2010)

[60] Nipper J et al. *Phys. Rev. X* **2** 031011 (2012)

[61] Tiarks D et al. *Phys. Rev. Lett.* **113** 053602 (2014)

[62] Kondo J M et al. *Phys. Rev. A* **90** 023413 (2014)

[63] Gorniaczyk H et al. arXiv:1511.09445v1 [quant-ph] (2015)

[64] Pelle B et al. arXiv:1510.05350v1 [physics.atom-ph] (2015)

[65] Beterov I I and Saffman M *Phys. Rev. A* **92** 042710 (2015)

[66] Bergmann K, Theuer H, and Shore B W *Rev. Mod. Phys.* **70** 1003 (1998)

[67] Raussendorf R and Briegel H *Phys. Rev. Lett.* **86** 5188 (2001)

[68] Jaynes E T and Cummings F W *Proc. IEEE* **51** 89 (1963)